\documentclass[12pt, preprint]{aastex}

\newcommand{\XMM}{{\em XMM-Newton }}
\newcommand{\Ch}{{\em Chandra }}
\newcommand{\ot}{[O~{\sc iii}] }

\def\gappeq{\mathrel{ \rlap{\raise.5ex\hbox{$>$}}
                      {\lower.5ex\hbox{$\sim$}}  } }
\def\lappeq{\mathrel{ \rlap{\raise.5ex\hbox{$<$}}
                      {\lower.5ex\hbox{$\sim$}}  } }

\shorttitle{MULTIWAVELENGTH OBSERVATIONS OF NGC 2110}
\shortauthors{EVANS ET AL.}

\begin{document}

\title{The Chandra, HST, and VLA View of the Circumnuclear Extended Emission in the Narrow Emission Line Galaxy NGC 2110}
\author{Daniel~A.~Evans\altaffilmark{1}, Julia~C.~Lee\altaffilmark{1}, Maria~Kamenetska\altaffilmark{2}, Sarah~C.~Gallagher\altaffilmark{3}, Ralph~P.~Kraft\altaffilmark{1}, Martin~J.~Hardcastle\altaffilmark{4}, and Kimberly~A.~Weaver\altaffilmark{5}}
\altaffiltext{1}{Harvard-Smithsonian Center for Astrophysics, 60 Garden Street, Cambridge, MA 02138}
\altaffiltext{2}{MIT Kavli Institute for Astrophysics and Space Research, 77 Massachusetts Avenue, NE 80, Cambridge, MA 02139}
\altaffiltext{3}{UCLA, Division of Astronomy \& Astrophysics, 430 Portola Plaza, Mail Code 154705, Los Angeles, CA 90095-1547}
\altaffiltext{4}{School of Physics, Astronomy \& Mathematics, University of Hertfordshire, College Lane, Hatfield AL10 9AB, UK}
\altaffiltext{5}{NASA Goddard Space Flight Center, Code 662, Greenbelt, MD 20771}

\begin{abstract}

We present results from new {\it Chandra} and archival {\it HST} and VLA imaging observations of the circumnuclear extended emission in the nearby Type 2 Seyfert galaxy NGC 2110. We find resolved soft-band X-ray emission $\sim4''$ ($\sim160$ pc) north of the nucleus, which is spatially coincident with \ot emission, but lies just beyond the northern edge of the radio jet in the source. We find that shock-heating of multi-phase gas clouds can successfully account for this extended emission, although we cannot rule out alternative models, such as the scattering of nuclear radiation by ionized material, or pure photoionization from the nucleus. In addition, we detect kpc-scale ($\sim30''$) extended soft-band X-ray emission south of the nucleus. Finally, we compare our results for NGC~2110 with the prototypical Type 2 Seyfert galaxy NGC~1068, and suggest that different physical processes could produce extended circumnuclear X-ray emission in Seyfert galaxies.

\end{abstract}


\section{INTRODUCTION}

X-ray observations of the circumnuclear environments of Seyfert galaxies with small-scale radio jets are important diagnostics for understanding the interplay between the ionizing radiation from the nucleus, radio ejecta, and galaxy ISM. In some of the brightest sources, e.g., NGC~1068 (e.g.,~\citealt{you01}), Mrk~3 (\citealt{sak00}), and the Circinus galaxy (\citealt{sam01}), high-resolution \Ch and \XMM grating spectroscopy reveals their X-ray spectra to be dominated by He-like emission lines most likely to be produced in a plasma photoionized by nuclear radiation. However, alternative models, such as shock-heating induced by the interaction of the radio jet and ambient medium may be applicable to other Seyfert galaxies. High resolution X-ray imaging observations, combined with multiwavelength data, are important probes of circumnuclear environments, and are complementary to X-ray grating data.

NGC 2110, the subject of this paper, is a nearby ($z=0.007579$, $D_{\rm L}=32.7$ Mpc) Seyfert type 2 galaxy (\citealt{shu80}). Historically, it is classified as a Narrow Emission Line Galaxy (NELG), which is the subclass of Seyfert galaxies with strong 2--10 keV X-ray emission and narrow ($<600$km s$^{-1}$) optical emission lines. The source exhibits complex morphology on a range of angular scales, and shows a rich variety of structure across the electromagnetic spectrum. VLA observations (\citealt{ulv89}) show that it hosts an `S'-shaped jet-counterjet system that extends $\sim4''$ ($\sim$600 pc) in the N-S direction, while HI spectral-line mapping (\citealt{gal99}) indicates the presence of a small-scale disk that may partially obscure the southern radio jet. Optically, NGC 2110 is classified as a S0 galaxy, although broad-band {\it HST} observations (\citealt{mul94,gon02,fer04}) have revealed a series of complex, spiral-like dust lanes with a distribution consistent with a tightly wrapped single-armed spiral.

Emission-line \ot$\lambda$5007 and H$\alpha$ + [N~{\sc ii}] {\it HST} observations of the inner regions of NGC 2110 (\citealt{mul94}) have revealed a variety of structures. The dominant line emission is concentrated into a strongly curved ``jet-like'' feature (as characterized by \citealt{mul94}) that extends $\sim1.4''$ ($\sim230$ pc) North of the nucleus and becomes increasingly broadened with distance. The \ot emission is noticeably more asymmetric about the nucleus than the H$\alpha$ emission, and \cite{mul94} suggest that this is at least in part due to a north-south gradient in excitation. Detailed spectroscopy of the ``jet-like'' feature with the {\it HST} Faint Object Spectrograph (FOS) (\citealt{fer99}) suggests that photoionizing-shock models (e.g.,~\citealt{ds96}) provide a good explanation of the origin of this emission, although photoionization by a central source cannot be ruled out. Recently, OASIS integral field spectroscopy of the same region (\citealt{fer04}) confirmed that the majority of the narrow \ot emission is not associated with the nucleus, and may rather be associated with the ``jet-like'' feature. This is in contrast to the H$\alpha$ emission, whose peak intensity is coincident with that of the radio nucleus. On slightly larger scales, \cite{mul94} also report the detection of an `S'-shaped distribution of ionized gas that extends $\sim4''$ in either direction of the nucleus. Although this feature, together with the ``jet-like'' feature, extend along a similar position angle to the radio emission, neither appears to be spatially coincident with the radio morphology. Instead, the strongest optical line emission occurs where the radio emission is weak, and the optical ``jet-like'' feature is more curved than the radio jet. This suggests that, although possibly influenced by the radio-emitting feature, the optical line emission is not directly associated with it.

At X-ray energies, {\it ROSAT} and {\it BBXRT} observations of NGC 2110 (\citealt{wea95}) showed the presence of spatially resolved X-ray emission that extends $\sim4''$ North of the unresolved nucleus, possibly coincident with the optical emission-line `S'-shaped feature. They argued that at least a fraction of the soft excess observed in the X-ray spectrum of NGC 2110 below 1~keV (e.g.,~\citealt{mus82}) originates in this extended region. However, the physical origin of the extended emission was unclear, and several models have been proposed to explain it. In one model, the impact of the radio jet into the ISM may induce shock-heating, producing the observed soft X-ray emission, together with the optical emission lines. \cite{fer99} provide an in-depth discussion of photoionizing shock models, and demonstrated that if {\it in situ} shock heating were to power the X-ray emission, shock velocities in excess of 500 km s$^{-1}$ would be required (c.f. the measured \ot linewidth of 200 km s$^{-1}$ in this region --- e.g.,~\citealt{wil85}). However, the shock-heating could occur in a region much closer to the nucleus, where the measured \ot velocities are higher, and subsequently expand outwards. Another interpretation, discussed by \cite{wea95}, is that the extended X-ray emission is electron-scattered nuclear radiation. Alternatively, the extended emission may result from pure photoionization of ambient gas by the nucleus. A final interpretation, given the common detection with \Ch of resolved kpc-scale X-ray jets in low-power radio galaxies (e.g.,~\citealt{wor01}), is that the resolved X-ray emission is from the VLA-detected radio ejecta. \cite{don04} showed a series of zeroth-order \Ch HETGS images of the nucleus of NGC~2110, which hinted at the presence of extended soft-band X-ray emission north of the nucleus, and we use the superb spatial resolution and good low-energy sensitivity of the \Ch ACIS instrument to provide further constraints on the nature of this emission.

In this paper, we present a multiwavelength imaging study of the circumnuclear extended emission in NGC~2110. We make a detailed comparison between the \Ch X-ray, and archival radio and optical observations of the extended emission north of the nucleus, and discuss models that might explain its origin. In addition, we present the first evidence for large-scale soft-band X-ray emission south of the nucleus. Finally, we compare our results for NGC~2110 with the prototypical Type 2 Seyfert galaxy NGC~1068, and argue that different physical processes could produce the extended circumnuclear X-ray emission in Seyfert galaxies. A detailed analysis of the \Ch HETGS spectra of NGC~2110 will be presented in a future paper (Lee et al., in prep.), and the energetics arguments presented in this paper are in part a prelude to our analysis of the high energy-resolution \Ch gratings data.

This paper is organized as follows. Section 2 contains a description of the data and a summary of their analysis. In Section 3, we present the \Ch imaging observation and compare it to observations with {\it HST} and the VLA. In Section 4, we discuss several models that may account for the small-scale northern extended emission, while in Section 5 we discuss the large-scale southern extended X-ray emission. In Section 6, we compare NGC~2110 with the prototypical Seyfert 2 galaxy NGC~1068. We end with our conclusions in Section 7. All results presented in this paper use a cosmology in which $\Omega_{\rm m, 0}$ = 0.3, $\Omega_{\rm \Lambda, 0}$ = 0.7, and H$_0$ = 70 km s$^{-1}$ Mpc$^{-1}$. At the redshift of NGC 2110\footnote{Taken from the SIMBAD database} ($z$ = 0.007579), 1 arcsec corresponds to 160 pc. This value is slightly different ($-131$ km s$^{-1}$) than the velocity (with respect to the CMB) derived from stellar Mg {\sc i}b absorption lines (\citealt{nel95}). However, none of our conclusions would be changed if we adopted this alternate value, and we note that previous publications on NGC~2110 have used a variety of redshift estimates. All spectral fits include absorption through our Galaxy using $N_{\rm H, Gal}$ = 1.76$\times$10\(^{21}\) cm$^{-2}$ (\citealt{dic90}). Errors quoted in this paper are 90 per cent confidence for one parameter of interest (i.e., $\chi^2_{\rm min}$ + 2.7), unless otherwise stated. When distinguishing between different model fits to the data, we consider $F$-statistic results, and we adopt a threshold of 95\%  for a significant improvement in the fitting statistic.

\section{OBSERVATIONS AND ANALYSIS}

NGC~2110 was observed on 2000 April 22 (OBSID 883) with the S3 chip of the {\it Chandra}/ACIS charge-coupled device (CCD) camera. The observation was made in the VFAINT mode, using a 128-row subarray (giving a $1 \times 8$ arcmin field of view), in order to reduce the CCD frame time to 0.44 s and lessen the effect of pileup. The source was positioned near the standard aimpoint on the S3 chip. The data were reprocessed using {\sc{CIAO v3.2.2}} with the {\sc{CALDB}} v3.10 calibration database to create a new level-2 events file filtered for the ASCA grades 0, 2, 3, 4, 6, and with the 0.5-arcsec pixel randomization removed. To check for intervals of high particle background, light curves were extracted for the ACIS-S3 chip, excluding the core. No periods of high background were found, and so the entire 44.3 ks observation was used in our analysis.

The \Ch observation of the nucleus of NGC~2110 is affected by pileup to the extent that the accurate modeling of the nuclear spectrum is significantly hampered. We demonstrate this using two independent methods. First, the counts extracted from a source-centered circle of radius 1.23 arcsec (2.5 pixels) in the 0.5--7 keV energy band gave 0.38 counts per frame, for which a pileup fraction of $\sim 15$--20 per cent is predicted using the {\sc{PIMMS}} software. As a second independent test, we estimated the spatial extent of the pileup by producing images consisting of grade 7 events, which are largely produced by photon pileup. Inspection of the images shows the pileup to be concentrated within the central 2 pixels.

In addition to the X-ray data, we utilize multiwavelength archival data from observations of NGC~2110 with the VLA and {\it HST}. NGC 2110 was observed with the VLA in its A configuration at 5.0 GHz for $\sim$70 minutes on 1994 February 27. We extracted the data from the archive and calibrated them in the standard manner with {\sc aips}, using 3C\,48 as the primary flux calibrator and J0541-056 as the phase calibrator. The quality of the final image of NGC~2110 was improved using one iteration of phase self-calibration, and the noise level (40 $\mu$Jy beam$^{-1}$) approaches the expected thermal limit. We verified by examining B-configuration archival data (taken on 1985 May 27) that the A-configuration data spatially samples all the detectable structure. The reduced {\it HST} data were provided to us by J. Mulchaey, and \cite{mul94} should be consulted for a detailed description of the data-reduction methods used.

\section{RESULTS}

\subsection{Soft-Band Extended Emission}
\label{softband}

\subsubsection{Small scale emission}

\begin{figure}
\begin{center}
\includegraphics[width=14cm]{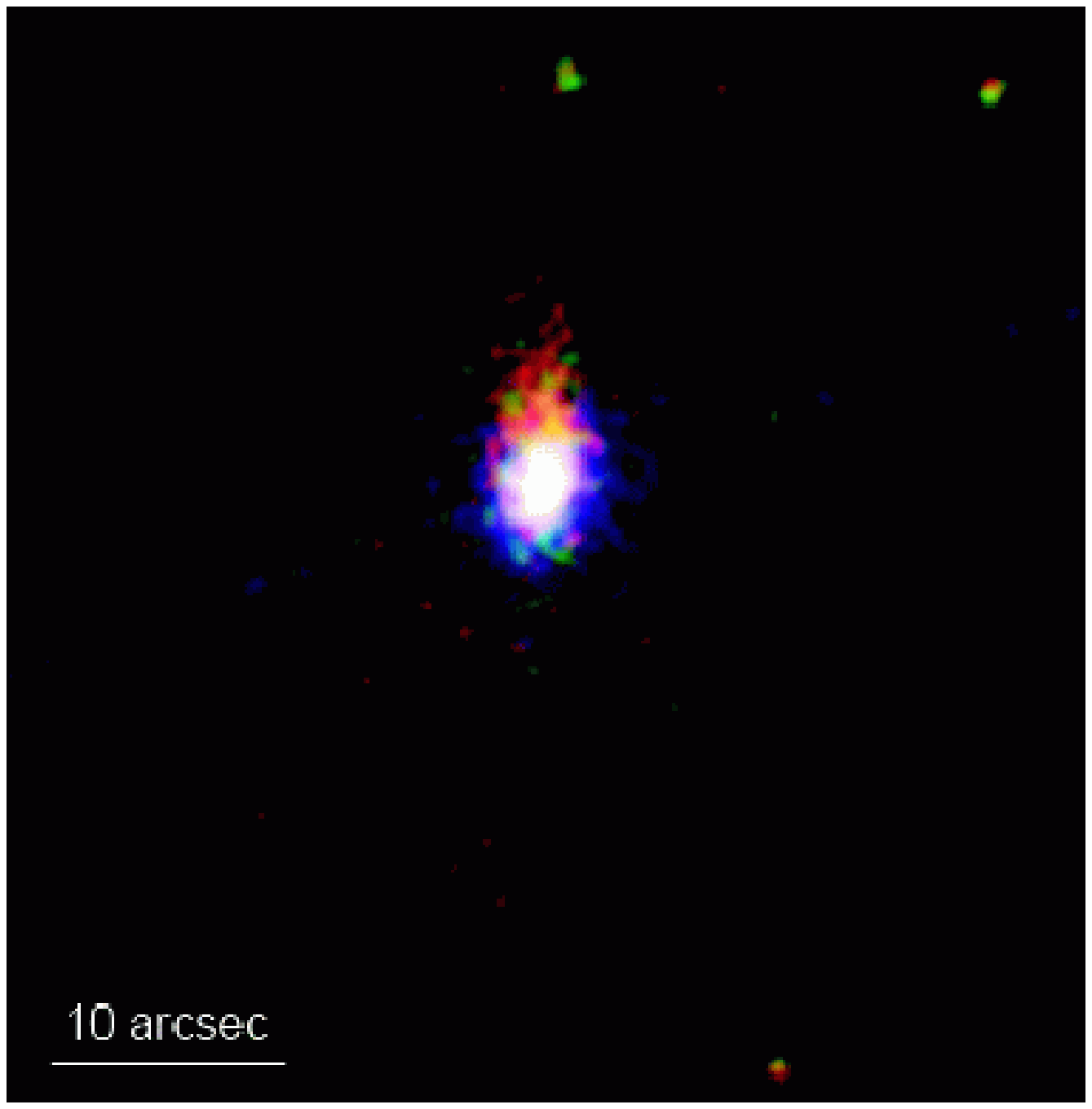}
\caption{Tri-color (red: 0.5--1 keV,  green: 1--1.5 keV, blue: 4--5~keV) ACIS-S image of NGC 2110, showing soft extended emission towards the north of the nucleus. The image has been binned to 0.25$''$ pixel$^{-1}$, and subsequently smoothed with a Gaussian of $\sigma=0.5''$.}
\label{colormap}
\end{center}
\end{figure}

Figure~\ref{colormap} shows a Gaussian-smoothed ($\sigma=0.5''$) tri-band \Ch ACIS-S image of NGC~2110. Here, red corresponds to 0.5--1 keV photons, green to 1--1.5 keV photons, and blue to 4--5 keV photons. We have deliberately emphasized the 0.5--1 keV photons, in order to contrast them with the 4--5 keV photons. Figure~\ref{colormap} shows that the extended X-ray emission to the north of the nucleus, first reported by \cite{wea95} with {\it ROSAT}, is dominated by soft X-ray photons. This is in contrast to the nucleus, which shows both soft and hard ($>$4 keV) X-ray emission. There is a hint of 1--1.5 keV X-ray emission just south of the nucleus, which may be emission similar to the northern feature, but obscured by the small-scale gas disk discussed by \cite{gal99}.

\begin{figure}
\begin{center}
\includegraphics[width=14cm]{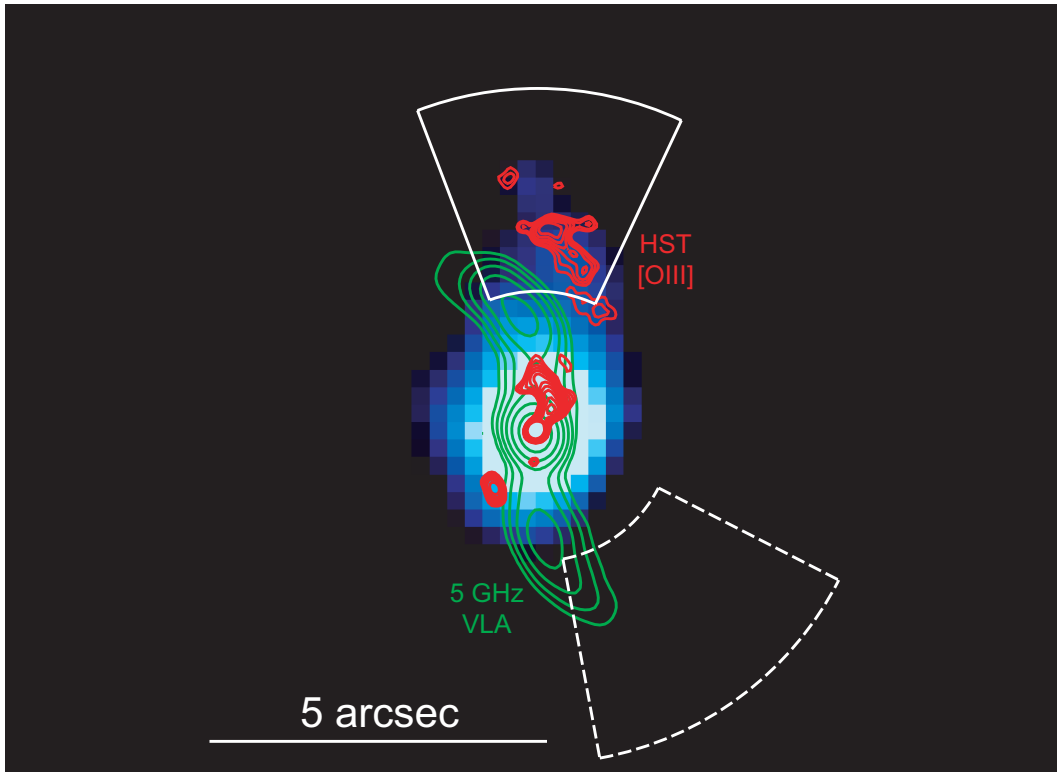}
\caption{0.5--1.5 keV \Ch image of NGC 2110 (blue) with VLA (green) and {\it HST} \ot (red) contours overlaid. The \Ch image has been binned to 0.25$''$ pixel$^{-1}$, and subsequently smoothed with a Gaussian of $\sigma=0.5''$. The relative astrometry of the X-ray, radio, and \ot images have been adjusted to make the unresolved cores spatially coincident. The white pie section marks the region used to extract the spectrum of the extended X-ray emission, and the dashed white section marks the background sampled.}
\label{multiwave}
\end{center}
\end{figure}

Figure~\ref{multiwave} shows a 0.5--1.5 keV \Ch image of NGC~2110, binned to 0.25$''$ pixel$^{-1}$ and subsequently smoothed with a Gaussian of $\sigma=0.5''$, with contours from {\it HST} \ot (\citealt{mul94}) and 5-GHz VLA radio observations overlaid. The relative astrometry of the X-ray, radio, and \ot images have been adjusted to make the unresolved cores spatially coincident. Figure~\ref{multiwave} shows that the X-ray and radio emission extend along a similar position angle north of the nucleus, but are not spatially coincident: the X-ray emission lies beyond the northern edge of the radio jet, before the jet begins to diverge away from it. However, the X-ray emission is coincident with the `S'-shaped \ot emission. Our results may support a common origin for the soft X-ray and \ot emission, which we discuss more fully in Section~\ref{n_interpretation}.

\begin{figure}
\begin{center}
\includegraphics[angle=270,width=14cm]{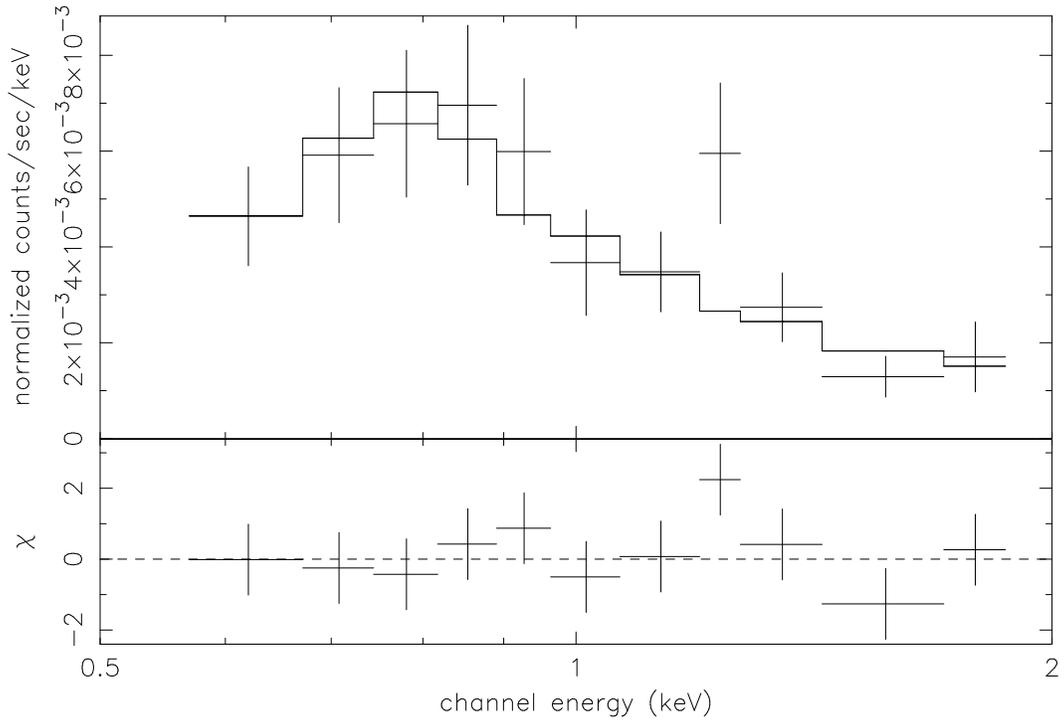}
\caption{0.5--2 keV X-ray counts spectrum and contributions to $\chi^2$ of the extension north of the nucleus. The model fit shown is the sum of two thermal components of temperature 0.34 and 5.5 keV, with abundance fixed at solar.}
\label{extensionspectrum_n}
\end{center}
\end{figure}

We extracted the X-ray spectrum of the extended X-ray emission, using a source-centered annulus of inner radius 2$''$ and outer radius 5$''$ between the position angles -25$^\circ$ and 20$^\circ$. In order to reduce any possible contamination from the wings of the unresolved nucleus, we adopted a matched background extraction region of identical radii between the position angles 190$^\circ$ and 235$^\circ$. The source and background extraction regions are marked in Figure~\ref{multiwave}. Imaging showed possible contamination from the nucleus at hard energies ($>$2 keV), and so we restricted our spectral analysis to 0.5--2 keV. The data were grouped to 20 counts per bin. The X-ray spectrum is shown in Figure~\ref{extensionspectrum_n}.

We attempted to fit a series of of models to the X-ray spectrum of the extended emission. Guided by the prevalence of jet-induced shock-heating in radio galaxies (e.g.~\citealt{kra03}), we initially fitted the X-ray spectrum of the extension with a single-temperature, collisionally ionized thermal plasma ({\sc Apec}) model, with abundance fixed at solar, and modified by Galactic absorption. However, the fit was poor ($\chi^2=35.4$ for 9 d.o.f.), due to the model greatly overproducing Fe L lines between 0.7 and 1.2 keV for the best-fitting $kT=0.62$ keV. The fit was significantly improved ($\chi^2=6.6$ for 8 d.o.f.) when the abundance was left to be a free parameter, although its value tended to an implausibly low value of $(1.01^{+3.23}_{-1.01})\times10^{-2}$ solar. An acceptable fit ($\chi^2=8.3$ for 7 d.o.f.) was achieved with the combination of two thermal models of abundances fixed at solar; one of temperature $kT_1=0.34^{+0.11}_{-0.07}$ keV, and the other significantly hotter, with a temperature $kT_2\sim5.5$ keV and normalization a factor of $\sim3$ higher than the cool component. Figure~\ref{extensionspectrum_n} shows the counts spectrum and $\chi^2$ residuals for this model. 

We fitted the X-ray spectrum with an alternative model consisting of an unabsorbed power law and multi-temperature blackbody emission from an accretion disk. This model is appropriate for the scattering of nuclear radiation by a population of electrons. The best-fitting nuclear spectrum is found from the \Ch HETGS data (Lee et al., in prep.). We emphasize that the intrinsic absorption of the nucleus ($N_{\rm H}=\sim4\times10^{22}$ cm$^{-2}$), which we associate with the central engine of NGC~2110, is not included in our model fit to the extended region. With the power-law photon index, disk temperature, and relative normalizations fixed at their HETGS values, we found an adequate fit to the extended-emission spectrum ($\chi^{2}=14.8$ for 10 d.o.f.), although there were noticeable negative residuals below $\sim0.8$ keV. Table~\ref{extensionspectrum_n_table} summarizes the results of our spectral fitting.

Finally, given the observations of photoionized line-emission in high energy-resolution X-ray observations of certain nearby Seyfert galaxies (e.g.,~\citealt{sak00,sam01,ogle03}), it is interesting to consider photoionized gas as the emission mechanism for the extended X-ray emission in NGC 2110. However, we note that the detection of photoionized line-emission is hampered here due to the relatively poor quality of the Chandra ACIS spectrum of the extended emission ($\sim$200 counts), together with its CCD energy resolution. In short, we cannot distinguish spectrally between these plausible models for the extended X-ray emission. However, the 0.5-2.0 keV X-ray luminosity of $\sim3\times10^{39}$ ergs s$^{-1}$ is relatively insensitive to the spectral modeling, and this constraint allows us to investigate simple arguments that demonstrate the energetic viability of photoionization (see Section~\ref{n_interpretation_photoionization}).

\subsubsection{Large scale emission}
\label{extension_s}

\begin{figure}
\begin{center}
\includegraphics[width=14cm]{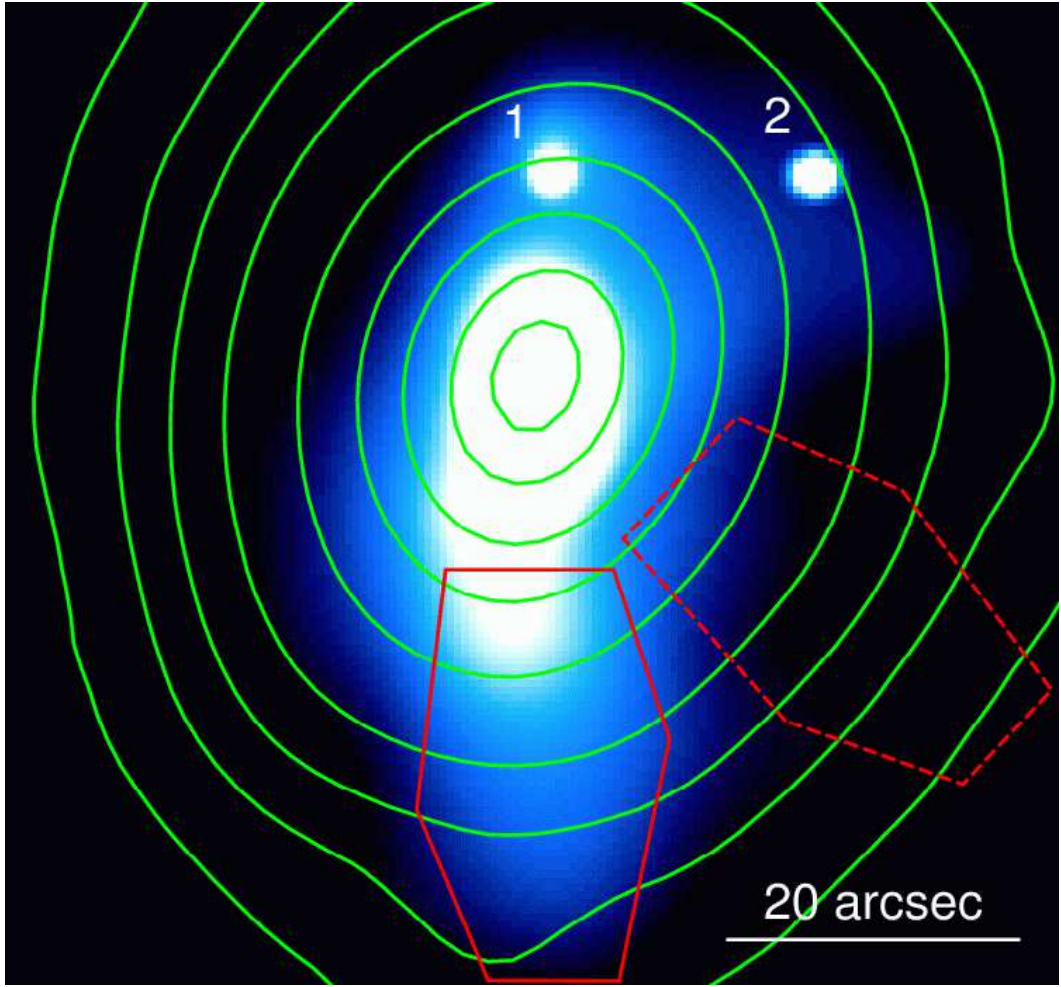}
\caption{Adaptively smoothed 0.5--1 keV \Ch ACIS-S image of NGC~2110, with optical DSS contours overlaid. The X-ray morphology does not trace the stellar population, with a noticeable X-ray extension south of the nucleus. Two point sources to the North of the nucleus, marked (1) and (2), are also observed. The large solid red box south of the nucleus marks the region used to extract the spectrum of the extended X-ray emission, and the dashed box marks the background extraction region.}
\label{adaptivesmooth}
\end{center}
\end{figure}

\begin{figure}
\begin{center}
\includegraphics[angle=270,width=14cm]{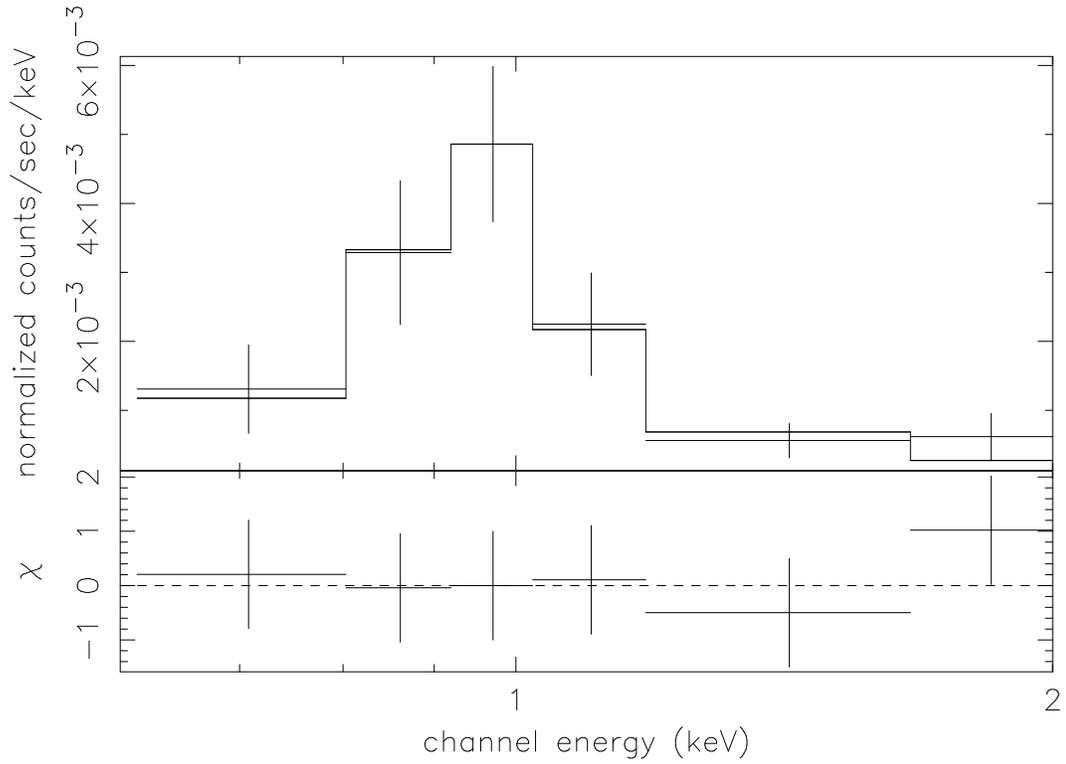}
\caption{0.5--2 keV X-ray counts spectrum and contributions to $\chi^2$ of the large-scale extension south of the nucleus. The model fit shown is the sum of a power law of photon index 1.6 and thermal emission of temperature 0.82 keV and abundance fixed at solar.}
\label{extensionspectrum_s}
\end{center}
\end{figure}

On larger ($\sim30''$) scales, there is a noticeable asymmetry between the soft X-ray emission located north and south of the nucleus. Figure~\ref{adaptivesmooth} shows an adaptively smoothed 0.5--1 keV image of NGC~2110, with contours from an optical DSS image overlaid. The X-ray morphology does not trace the stellar population. We extracted the 0.5--2 keV spectrum of the large-scale southern extension from the source and background regions marked in red in Figure~\ref{adaptivesmooth}. The spectrum, grouped to 20 counts per bin, is shown in Figure~\ref{extensionspectrum_s}. A model fit to a single-component thermal {\sc Apec} model of temperature $0.96^{+0.21}_{-0.17}$ keV and abundance fixed at solar provided a good fit to the spectrum ($\chi^2=1.4$ for 4 d.o.f.). The parameters of the best-fitting model are shown in Table~\ref{extensionspectrum_s_table}.

\subsection{Associated Point Sources?}
\label{sectionpointsources}

Figure~\ref{adaptivesmooth} also shows two point sources to the North of the nucleus of NGC~2110. The sources, marked (1) and (2), are most prominent at energies $\lappeq$1.5 keV. We inspected an archival {\it HST} continuum image and found no optical counterparts to these sources. There are 61 and 74 background-subtracted 0.5--7 keV counts in each, respectively. Their 0.5--10 keV X-ray spectra are well described by unabsorbed power laws of photon index 1.6$\pm$0.4 and 1.2$\pm$0.3, respectively, with unabsorbed fluxes $(1.4\pm0.6)\times10^{-14}$ and $(2.2\pm0.5)\times10^{-14}$ ergs s$^{-1}$ cm$^{-2}$. We performed queries with the NED and SIMBAD databases, and found no known sources in their vicinity. We used the soft-band $\log N-\log S$ relationship for the \Ch Deep Fields (\citealt{toz01,bau04}) to estimate the number of background X-ray sources expected within a circle of radius $50''$ centered on the nucleus, and found this number to be $\sim0.5-1.5$, which is approximately consistent with the two sources we detected. While we cannot determine the distance to these sources, if they were at the redshift of NGC~2110, their 0.5--10 keV unabsorbed luminosities would be $(1.6\pm0.5)\times10^{39}$ and $(2.7\pm0.7)\times10^{39}$ ergs s$^{-1}$, respectively, which would make them Ultraluminous X-ray Sources (ULXs) (e.g.,~\citealt{swa04}).

\subsection{Hard-band X-ray emission}

\begin{figure}
\begin{center}
\includegraphics[width=14cm]{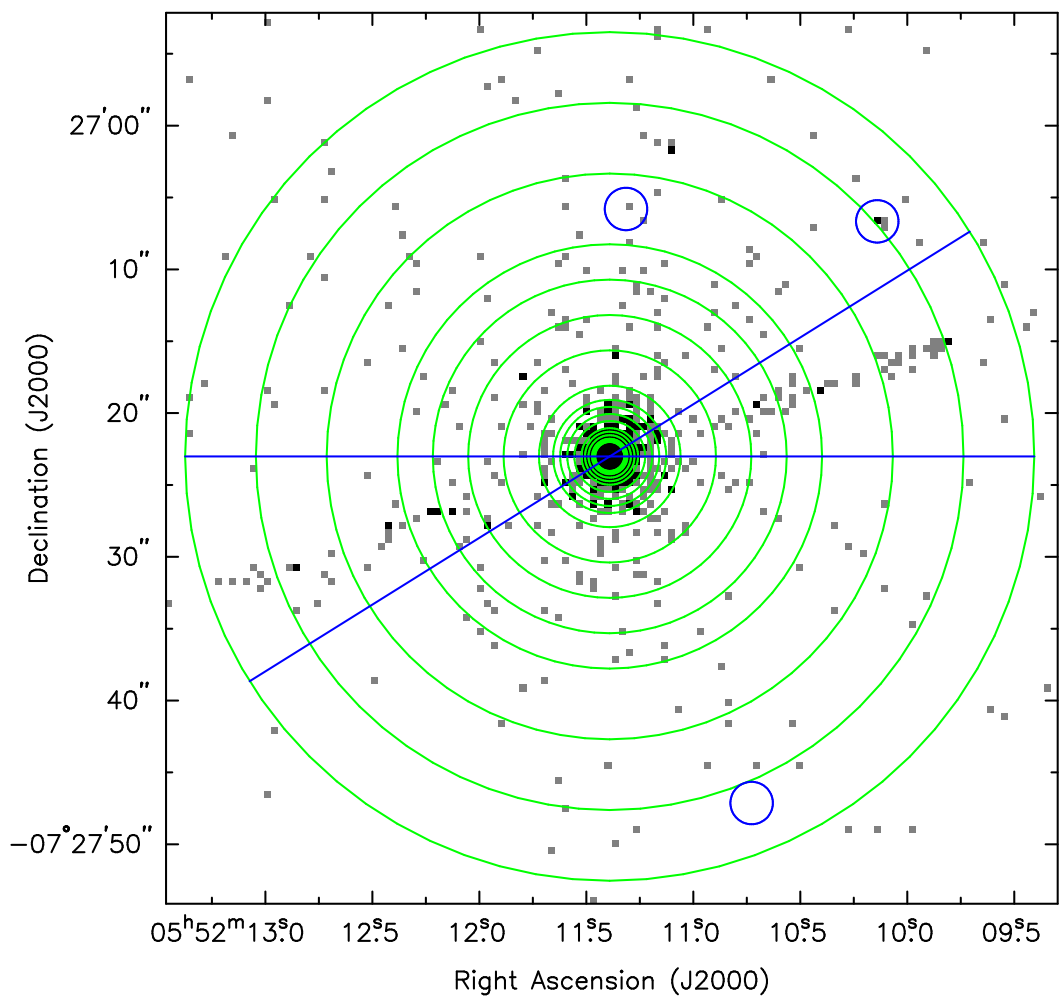}
\caption{4--4.5 keV \Ch image of NGC~2110, showing the regions used to extract the radial profile. The radial profile, sampled from the concentric annuli shown in green, extends from 1$''$  to 30$''$, with background taken from a large source-centered annulus of inner radius 37$''$ and outer radius 50$''$. The profile excludes pie slices between the position angles 90$^\circ$--122$^\circ$ and 270$^\circ$--302$^\circ$, in order to exclude emission from the frame transfer streak, and circles that mask emission from non-nuclear point sources.}
\label{4-4.5_rp_regions}
\end{center}
\end{figure}

\begin{figure}
\begin{center}
\includegraphics[width=8cm]{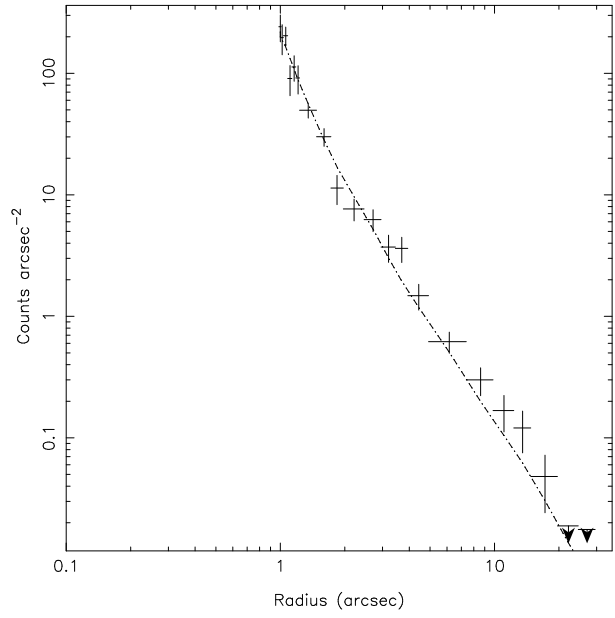}
\caption{Background-subtracted 4--4.5 keV radial surface brightness profile of NGC~2110, shown with a model \Ch PSF.}
\label{4-4.5_rp}
\end{center}
\end{figure}

We attempted to confirm that the hard ($>$3 keV) X-ray emission is unresolved by extracting and modeling the radial profile of NGC~2110 from the ACIS-S3 chip. The results described here are for fits to 4--4.5 keV events, but are consistent with the results for any $>$3 keV events. The 4--4.5 keV radial profile was centered on the nucleus, and excluded pie slices between the position angles 90$^\circ$--122$^\circ$ and 270$^\circ$--302$^\circ$, in order to exclude emission from the frame transfer streak, and circles that mask any emission from the point sources discussed in Section~\ref{sectionpointsources}. The radial profile extended from 1$''$ (so as to exclude events known to be affected by photon pileup), out to 30$''$, with background taken from a large source-centered annulus of inner radius 37$''$ and outer radius 50$''$. Figure~\ref{4-4.5_rp_regions} shows the regions used to extract the radial profile. A PSF of $\Gamma=1.6$ was modeled using ChaRT and MARX, and was convolved with a Gaussian of r.m.s. width 0.35$''$ in order to provide a good match with the radial profile data close to the core. The radial profile is shown in Figure~\ref{4-4.5_rp}.

An acceptable fit to the radial profile ($\chi^{2}=21.2$ for 20 dof) was achieved with the model of the \Ch PSF. However, at distances $\gappeq8''$, there is a small excess of observed counts with respect to the model PSF. We inspected the 4--4.5 keV image, and found a slight excess of counts towards the northwest of the nucleus. While the origin of the excess emission is unclear, it is not spatially coincident with the soft X-ray emission, which has a north-south orientation. We speculate that, given the readout direction of the ACIS instrument, the apparent excess hard emission is due to the combination of serial and parallel charge transfer inefficiency (M. Nowak, private communication).

\section{ORIGIN OF NORTHERN EXTENDED EMISSION}
\label{n_interpretation}

\subsection{X-ray emission from the radio jet?}
\label{n_interpretation_jet}

Many low-power radio galaxies show X-ray jets that are inferred to have a synchrotron origin, given the good spatial agreement between the X-ray and radio emission. Since the X-ray and radio emission in NGC~2110 are not spatially coincident (Fig.~\ref{multiwave}), it seems unlikely that the X-ray emission is directly associated with the radio jet.

\subsection{Shock-heated gas?}
\label{n_interpretation_shock}

An alternative model is that clouds of gas have been ablated and shock-heated by the radio jet. We test this model by comparing the pressures of the radio jet with the X-ray and \ot emission, in order to determine if the pressure of the jet is sufficient to drive a shock into the gas. We adopt the standard assumption of equipartition between the magnetic field and particles to estimate the minimum pressure of the radio jet. The 5-GHz flux density of the brightest part of the jet is 40 mJy. We model the jet as an ellipsoid of dimensions $1.4''\times0.2''\times0.2''$ (volume $3.5\times10^{60}$ cm$^{3}$). We assume the electron energy distribution extends from a Lorentz factor $\gamma_{\rm min}=2$ up to $\gamma_{\rm max}=10^{5}$, with an electron energy index $p=2.4$ (the results have only a weak dependence on the choice of $\gamma$ - see \citealt{har04}). Under the simple assumption that the jet is in the plane of the sky, and that relativistic beaming is unimportant, the minimum pressure is $\sim10^{-10}$ Pa, and the minimum energy is $\sim10^{52}$ ergs.

In order to estimate the pressure of the hot gas, we adopt a two-temperature plasma as the best-fitting spectral model (Model N2 in Table~\ref{extensionspectrum_n_table}). We adopt a cylindrical region, of radius 0.37$''$ and length 1.9$''$ (volume $9.9\times10^{61}$ cm$^{3}$), located where the X-ray emission is brightest and is adjacent to the radio emission. The {\sc Apec} normalization is $10^{-14}(1+z)^2\int n_e n_p dV/4\pi D^2_L$, where $n_e$ and $n_p$ are the electron and proton number densities (cm$^{-3}$), and $D_L$ is the luminosity distance (cm). From this, we find the pressure of the gas ($P\sim2.2n_p k_B T$) to be $\sim3\times10^{-9}$ Pa, and its energy $\sim3\times10^{54}$ ergs. We calculate the sound speed of the gas to be $\sim8\times10^{5}$ m s$^{-1}$, and find the characteristic sound crossing time to be $\sim2\times10^{5}$ years. The rate of energy transfer from the jet is therefore $\sim5\times10^{41}$ ergs s$^{-1}$. Finally, \cite{fer99} used {\it HST} FOC observations to estimate the temperature and density of the optical line-emitting gas. They derive an \ot temperature of $\sim$17,400 K and, using [S~{\sc ii}] line ratios, find a density $\sim1200$ cm$^{-3}$, implying a pressure $\sim6\times10^{-10}$ Pa.

The pressure of the X-ray gas is approximately 30 times greater than the minimum internal pressure in the radio jet. However, for a shock to be driven into the gaseous atmosphere, the pressure of the radio jet must be comparable to that of the X-ray gas. Although the radio pressure calculated is formally a {\it minimum} pressure, and the underlying assumption that the energies in radiating electrons and magnetic field are close to equipartition may not be applicable in the jet of NGC~2110. As an example, the diffuse radio structures that are apparent in FRI-type radio galaxies require large departures from equipartition in order to maintain pressure balance with the gaseous atmosphere (e.g.,~\citealt{cro03}), and we speculate that a similar situation may be occurring in NGC~2110. Indeed, we suggest that departures from equipartition might be an important difference between the `frustrated', diffuse jets and lobes observed in Seyfert galaxies such as NGC~2110 and FRI-type radio galaxies, and the powerful jets observed in FRII-type radio galaxies radio galaxies.

The spatial offset between the radio, and \ot and X-ray emission can be accounted for in this simple model of shock heating. As has been observed in the radio galaxy Centaurus~A (e.g.,~\citealt{kra03}), radio ejecta can be driven into a gaseous environment, giving rise to a cap of X-ray-emitting shock-heated gas that surrounds the radio emission, but is spatially distinct from it. We suggest that a similar mechanism may be applicable to NGC~2110, where the shock is detached.

An alternative possibility is that there is a considerable ram velocity of the jet into the gas, i.e., that the observed X-ray--emitting gas is already present and has not been shock-heated. This obviates the need for a balance between the ambient pressure of the gas and the internal pressure of the jet. However, this possibility has two problems. First, the origin of the X-ray--emitting gas would again be unclear, and second, the similarities in the spatial distribution of radio and X-ray gas would be merely coincidental.

A final interpretation within the shocked-gas scenario is that EUV photons from the bow shock of the radio lobe are sufficient to photoionize the ambient environment (e.g.,~\citealt{ds96}), giving rise to \ot and X-ray emission, as discussed by, e.g.,~\cite{fer99}. However, this model is highly unlikely to be able to generate X-ray emitting gas with temperatures $\gappeq0.1$ keV, since the emissivity of the required strong shock is extremely low.

We conclude that a simple model of shock-heated multi-phase gas clouds is an energetically viable mechanism for producing the observed optical emission lines and two-temperature X-ray emission.

\subsection{Scattered nuclear radiation?}
\label{n_interpretation_scattered}

A second possible scenario explaining the extended soft X-ray emission is that nuclear radiation is being scattered by a population of electrons. We evaluate this model by presenting simple energetic arguments based on the comparison of the scattered and direct nuclear fluxes, following \cite{wea95}.

We use \Ch HETGS observations (Lee et al., in prep.) to model the direct nuclear emission as the sum of a power law and multiple-component disk blackbody emission, modified by an absorbing column $N_{\rm H}\sim4\times10^{22}$ cm$^{-2}$. We adopt the best-fitting spectral model of the extended X-ray emission to consist of the same spectral components, but different normalizations, and without the intrinsic absorption (Model N3 in Table~\ref{extensionspectrum_n_table}).

The ratio of the unabsorbed power-law fluxes of the scattered and `direct' components is $\sim10^{-4}$. We assume the nuclear radiation to be emitted isotropically, and scattered into a cone of opening angle 45$^\circ$ ($\Omega=0.48$). The optical depth to Compton scattering is then $\sim5\times10^{-3}$ ($N_{\rm{scat}}\sim10^{22}$ cm$^{-2}$). Assuming the line-of-sight scattering radius to be $\sim240$ pc, we find the number density of electrons to be $\sim10$ cm$^{-3}$. This gives a mass of $\sim10^{7}$ M$_\odot$, which seems very large.

It is therefore unlikely that the scattering of isotropic nuclear radiation can power the extended X-ray emission. On the other hand, if the nuclear radiation were emitted anisotropically into a narrow cone, it has been argued by \cite{wea95} that electron-scattering may remain a viable mechanism, and our findings may be consistent with this interpretation. However, modeling of the mid-infrared SED of NGC~2110 (e.g.,~\citealt{fri06}) suggests that the dust in the `torus' is heated essentially isotropically by the optical--UV continuum, and we see no reason for a departure from isotropic emission at X-ray wavelengths. Finally, we note that a test of this model would be to observe polarized optical light in this region.

\subsection{Photoionization}
\label{n_interpretation_photoionization}

In this scenario, pure photoionization of gas by the nuclear radiation is responsible for the extended soft X-ray emission. The detection of photoionized line-emission is hampered in our relatively low signal-to-noise, CCD-quality spectrum, though we can still use the superior spatial resolution of \Ch to provide new constraints on the energetic viability of this model, using the measured X-ray luminosity of the emission, and the simple energetics arguments first proposed by \cite{wea95}. The luminosity in emission lines for a sphere of radius $r$ is 

\begin{equation}
L_{\rm{lines}}=\frac{4\pi}{3} n^2r^3j(\xi)
\end{equation}

\noindent where $n$ is the electron number density (cm$^{-3}$), $j(\xi)$ is the emissivity of the photoionized gas (ergs cm$^{3}$ s$^{-1}$) and $\xi$ is the ionization parameter (ergs cm s$^{-1}$) (see \citealt{wea95} for discussions). Assuming a sphere of radius 1.5$''$, and using the integrated 0.5--2 keV unabsorbed X-ray luminosity of the extended region of $\sim3\times10^{39}$ ergs s$^{-1}$, we find a number density of 1.35 cm$^{-3}$. For our calculations, we use the definition of the ionization parameter

\begin{equation}
\xi=\frac{L_{\rm{s}}}{nR^2}
\end{equation}

\noindent where $L_{\rm{s}}$ is the luminosity of the nucleus and $R$ is the distance from the nucleus to the photoionized gas. For $\xi\sim100$ ergs cm s$^{-1}$ (see arguments by \citealt{wea95}) and a distance $R=3.3''$, the luminosity of the nucleus required to photoionize the gas is in excess of $10^{44}$ ergs s$^{-1}$. This is a factor $\sim20$ greater than the unabsorbed nuclear luminosity. We conclude that photoionization is a viable mechanism for producing the extended emission (or at least a fraction of it), although as first noted by~\cite{wea95} the nuclear radiation may have to be emitted anisotropically if the extended emission were attributed solely to photoionization.

We note that the electron number density we calculate for the photoionized gas is approximately consistent with that which we derived for the collisionally ionized case (Section~\ref{n_interpretation_shock}). This demonstrates that we are not orders of magnitude off in our derivation of the properties of the photoionized gas, despite our simple assumptions.

Finally, if the extended soft X-ray emission were gas photoionized by the nucleus, then one would expect to observe emission lines from photoionized plasma in high signal-to-noise, high-resolution X-ray spectra. Observations with \Ch HETGS in AO-3 and AO-4 (Lee et al., in prep.) will help test this further, although even these data are not sufficient in terms of numbers of counts to permit a spectral analysis of the extended region. Preliminary results indicate that the soft-band nuclear X-ray emission is not dominated by photoionized emission lines, unlike that of, e.g., NGC~1068 (\citealt{ogle03}). The emission is instead consistent with direct emission from the nucleus that intersects cold obscuring material, possibly in the form of a torus. However, we tentatively detect H-like O~VIII, and He-like species of Ne~IX and O~VII, which may be consistent with photionization. The energetics arguments in this section simply serve to demonstrate that that photoionization is energetically viable.

\section{Origin of southern extended emission}

The extended X-ray emission to the south of the nucleus is interestingly not associated with any radio, optical-line, or CO emission, making its interpretation difficult. We investigate several models for the extended emission here.

First, we calculate the energy of the extended emission, by modeling the emission region as a cylinder of radius $6''$ and length $30''$ (volume = $4\times10^{65}$ cm$^{3}$). Using the same arguments as in Section~\ref{n_interpretation_shock}, we find the energy of the gas to be $10^{55}$ ergs. We speculate that if the extended X-ray emission may be related to the nucleus in some form of outflow.

Alternatively, some Seyfert galaxies show extended X-ray emission on scales of several kpc, coincident with an extended narrow line region (ENLR) (e.g.~\citealt{bia06}). However, owing to the generally poor signal-to-noise observations, the X-ray emission mechanism is uncertain, except for certain nearby Seyfert galaxies in which the extended emission is associated with photoionization. In NGC~2110, no ENLR is detected either to the North or South of the nucleus.

A final possibility is that the X-ray emission is similar to the ``anomalous arms'' seen in a \Ch observation of NGC~4258 (\citealt{wil01}). For NGC~4258, the X-ray emission is associated with gas from an inclined ($i=64^{\circ}$) disk, post-shocked by gas driven down onto the disk by the radio ejecta. For NGC~2110, this model does not explain why only X-ray emission is observed, rather than being accompanied by the H$\alpha$ emission seen in NGC~4258. Further, the large (projected) difference in the sizes of the radio and X-ray emission are not easily explained unless the galaxy is highly inclined, contrary to observations.

We conclude that the origin of this extended X-ray emission is unclear. However, its prominence to the south of the nucleus is consistent with its being on the portion of the disk nearest the observer, though we cannot definitively determine whether or not it is obscured by the disk.

\section{Comparison with NGC~1068}

NGC~1068 is a nearby ($z=0.003793$, $D_L=16.3$ Mpc) Sb spiral galaxy, and is often regarded as the prototypical example of a Seyfert 2 nucleus. An imaging observation with \Ch (\citealt{you01}) shows that the X-ray morphology of the source consists of a bright, compact ($\sim170$ pc) central region, a ($\sim550$ pc) extended region that points to the northeast, and a further (several kpc) region that appears to be associated with the spiral arms of the galaxy. The $\sim550$ pc NE X-ray region in NGC~1068 is spatially coincident with both the \ot emission {\it and} radio ejecta, unlike in NGC~2110, in which the X-ray and radio emission are spatially distinct. This may be evidence that in NGC~2110 the interaction of the radio jet and its environment produces a cap of shock-heated gas, unlike in NGC~1068. \cite{you01} argue that the northeastern region in NGC~1068 is not associated with starburst activity. In addition, \cite{you01} find that a single, collisionally ionized, thermal plasma model provides a poor fit to the spectrum of this region, unless the abundance tends to an implausibly low value. They instead fit the spectrum with the combination of two bremsstrahlung continua, of temperature $\sim$0.4 and 2.8 keV, plus a series of emission lines, with line energies consistent with results from \XMM RGS observations (\citealt{kin02}). Detailed modeling of the high-resolution X-ray spectrum of NGC~1068 (\citealt{bri02,kin02,ogle03}) shows that the emission lines are likely to be associated with a photoionized plasma. This is consistent with UV results from {\it HST} STIS emission-line mapping observations (\citealt{gro04}). NGC~1068 additionally shows evidence of a region of optical line-emitting gas at a distance $\sim200$ pc from the nucleus, which \cite{krae00} interpret as having been {\it locally} enhanced by the interaction of emission-line knots with the lower density interstellar medium. Although it is plausible that a similar process may produce the optical line-emission in NGC~2110, we consider it unlikely that it is able to produce the high-temperature X-ray gas observed, since the radiative coupling of $\sim$keV gas is extremely low. In summary, although the relatively poor quality of the CCD spectrum of the extended emission in NGC~2110 does not enable us to detect lines that might be consistent with emission from a photoionized plasma, the differing relative spatial distributions of X-ray, radio, and \ot emission in NGC~2110 and NGC~1068 may suggest that different physical processes could produce the extended circumnuclear X-ray emission in Seyfert galaxies. This is important to keep in mind when interpreting spatially unresolved X-ray data of distant Seyfert galaxies, e.g., those in deep fields.

\section{CONCLUSIONS}

We have presented results from a \Ch observation of the circumnuclear environment of the Seyfert 2 galaxy NGC~2110, and have combined them with VLA radio and {\it HST} \ot data, in order to constrain the physical processes present. To summarize our conclusions:

\begin{enumerate}

\item We detect resolved, soft-band X-ray emission $\sim4''$ North of the nucleus. The X-ray emission is spatially coincident with \ot optical line-emission, but not with the radio emission, and instead appears to lie north of the radio ejecta.

\item We consider several different models that may account for the extended northern X-ray emission. We find that it is possible that the radio jet has driven a strong shock through a series of multi-phase clouds of gas. We demonstrate that nuclear radiation scattered by a population of electrons may account for the extended X-ray emission, if the nucleus emits its radiation anisotropically. Finally, we show that photoionizaton of ambient gas by the nucleus may be sufficient to be responsible for the extended emission; high signal-to-noise \Ch grating observations can directly test this model. Though energetically all of these mechanisms are plausible, we consider shock emission from the radio jet interacting with the ISM to be the favored model because of the spatial properties of the multiwavelength emission.

\item We detect large-scale ($\sim30''$) extended X-ray emission South of the nucleus in NGC~2110, which we model as a collisionally ionized plasma of temperature $\sim0.9$ keV. The origin of this emission is unclear, although its prominence on the Southern side is consistent with the overall geometry of the galaxy.

\item A comparison of the spatial distributions of X-ray, radio, and \ot emission of NGC~2110 and the prototypical Seyfert 2 galaxy NGC~1068 leads us to suggest that different physical processes could produce the extended circumnuclear X-ray environments in Seyfert galaxies.

\end{enumerate}

\acknowledgements

This work was supported from the National Aeronautics and Space Administration, through a grant issued by the Chandra X-ray Observatory Center, which is operated by the Smithsonian Astrophysical Observatory for and on behalf of the National Aeronautics and Space Administration under contract NAS8-03060. MJH thanks the Royal Society for a Research Fellowship. Support for SCG was provided by NASA through the {\it Spitzer} Fellowship Program, under award 1256317. We thank John Mulchaey for providing us his reduced {\it HST} data, Fred Dulwich for assistance with {\sc AIPS}, and Andy Young, Chris Reynolds, and Mike Nowak for useful discussions. The National Radio Astronomy Observatory is a facility of the National Science Foundation operated under cooperative agreement by Associated Universities, Inc. This research has made use of the NASA/IPAC Extragalactic Database (NED) which is operated by the Jet Propulsion Laboratory, California Institute of Technology, under contract with the National Aeronautics and Space Administration, and of the SIMBAD database, operated at CDS, Strasbourg, France.

\begin{deluxetable}{lllc}
\tablecaption{Parameters of model fits to the 0.5--2 keV X-ray emission north of the nucleus}
\tablehead{
Model & Description & Parameters                & $\chi^2$/d.o.f. }
\startdata
N1    & TH                & $kT=0.62\pm0.05$                & 35.4/9          \\
      &                   & $Z=1$ (f)                                       \\ 
      &                   & norm$=(7.55\pm1.37)\times10^{-6}$                 \\
N2    & TH+TH             & $kT_1=0.34^{+0.11}_{-0.07}$ keV & 8.31/7          \\
      &                   & $Z_1=1$ (f)                                       \\
      &                   & norm$_1$$=(7.05^{+2.50}_{-2.86})\times10^{-6}$    \\
      &                   & $kT_2=5.5^{+59}_{-2.9}$ keV                                 \\
      &                   & $Z_2=1$ (f)                                       \\
      &                   & norm$_2$$=(1.98^{+2.09}_{-0.65})\times10^{-5}$    \\
N3    & DISKBB+PL         & $kT=0.19$ keV (f)               & 14.8/10         \\
      &                   & norm$_{\rm{diskbb}}=2.39\pm0.32$                  \\
      &                   & $\Gamma=1.6$ (f)                                  \\
      &                   & norm$_{\rm PL}$$=1.72\times10^{-6}$ (t) \\
\enddata
\label{extensionspectrum_n_table}
\tablecomments{Col. (2): TH=Collisionally ionized plasma ({\sc Apec}) model, DISKBB=Disk blackbody, PL=Power Law. Col. (3): (f) indicates parameter was frozen, (t) indicates parameter was tied to blackbody normalization. Normalization quoted is in units of $10^{-14}(1+z)^2\int n_e n_p dV/4\pi D^2_L$ (thermal), $[(R_{\sc in}/{\sc km})/(D_L/10 {\sc kpc})]^2\cos(i)$ (disk blackbody), and photons cm$^{-1}$ s$^{-1}$ keV$^{-1}$ at 1 keV (power law).}
\end{deluxetable}

\begin{deluxetable}{lllc}
\tablecaption{Parameters of model fits to the 0.5--2 keV X-ray emission south of the nucleus}
\tablehead{
Model & Description & Parameters                & $\chi^2$/d.o.f. }
\startdata
S1    & TH                & $kT=0.96^{+0.21}_{-0.17}$ keV & 1.4/4       \\
      &                   & $Z=1$ (f)                           \\
      &                   & norm$(4.23^{+1.25}_{-1.21})\times10^{-6}$  \\

\enddata
\label{extensionspectrum_s_table}
\end{deluxetable}

\end{document}